\documentclass{aa}
\usepackage{graphicx,natbib}
\bibpunct{(}{)}{;}{a}{}{,}

\def\apj{ApJ}

\def\araa{ARAA}

\def\flux{\rm erg~ s$^{-1}$~ cm$^{-2}$}
\def\lum{\rm erg~ s$^{-1}$}

\begin{document}

\sloppypar

   \title{X-ray emission from the stellar population in M32}

   \author{M.~Revnivtsev \inst{1,2}, E.~Churazov \inst{1,2}, S.~Sazonov
 \inst{1,2}, W.~Forman \inst{3}, C.~Jones\inst{3}}

   \offprints{mikej@mpa-garching.mpg.de}

   \institute{
              Max-Planck-Institute f\"ur Astrophysik,
              Karl-Schwarzschild-Str. 1, D-85740 Garching bei M\"unchen,
              Germany,
      \and
              Space Research Institute, Russian Academy of Sciences,
              Profsoyuznaya 84/32, 117997 Moscow, Russia
       \and
               Harvard-Smithsonian Center for Astrophysics, 
               60 Garden Street, Cambridge, MA 02138, USA
            }
  \date{}

        \authorrunning{Revnivtsev et al.}
        %\titlerunning{}

\abstract{Using {\sl Chandra} observations, we study the X-ray emission of
the stellar population in the compact dwarf elliptical galaxy M32. The
proximity of M32 allows one to resolve all bright point sources with
luminosities higher than $8\times10^{33}~ {\rm erg~s^{-1}}$ in the
0.5--7 keV band. The remaining (unresolved) emission closely follows
the galaxy's optical light and is characterized by an emissivity per
unit stellar mass of $\sim 4.3\times10^{27} {\rm erg~s^{-1}}
M_\odot^{-1}$ in the 2--10 keV energy band.  The spectrum of the
unresolved emission above a few keV smoothly joins the X-ray
spectrum of the Milky Way's ridge measured with {\sl RXTE} and
{\sl INTEGRAL}. These results strongly suggest that weak discrete X-ray
sources (accreting white dwarfs and active binary stars) provide the bulk
of the ``diffuse'' emission of this gas-poor galaxy. Within the
uncertainties, the average X-ray properties of the M32 stars are
consistent with those of the old stellar population in the Milky
Way. The inferred cumulative soft X-ray (0.5--2~keV) emissivity is however
smaller than is measured in the immediate Solar vicinity in our
Galaxy. This difference is probably linked to the contribution of 
young (age $\ll 1$~Gyr) stars, which are abundant in the Solar
neighborhood but practically absent in M32. Combining {\sl Chandra}, {\sl RXTE}
and {\sl INTEGRAL} data, we obtain a broad-band (0.5--60 keV) X-ray spectrum of
the old stellar population in galaxies. 
\keywords{ISM: general -- Galaxies:dwarf -- Galaxies:individual (M32) -- Galaxies: general -- Galaxies: stellar conent -- X-rays:diffuse background }}

   \maketitle

%
%________________________________________________________________

\section{Introduction}
Normal galaxies (i.e. those lacking a bright active nucleus) are often
observed as X-ray bright objects. In spiral galaxies, the main
contributors to the X-ray emission are the populations of low- and
high-mass X-ray binaries (LMXBs and HMXBs, see
e.g. \citealt{fabbiano03,gilfanov04} for reviews). Elliptical galaxies
lack HMXBs associated with recent star formation episodes, but
instead contain large quantities of hot ($\sim 0.5$ keV)
interstellar gas \cite[e.g.][]{forman85,canizares87}. The amount of
hot gas correlates with the galaxy mass, although with considerable
scatter, and in massive systems the hot gas provides the bulk of the
X-ray emission.

In the Milky Way, apart from the emission from LMXBs and HMXBs an
additional component is clearly seen, spatially concentrated to the
Galactic plane -- the so-called Galactic ridge X-ray emission (GRXE,
e.g. \citealt{worrall82}). The luminosity of this component amounts to
only a few per cent of the total X-ray luminosity of the Milky Way,
dominated by bright LMXBs and HMXBs.  The origin of this component
remained controversial for two decades, until recently a convincing
body of evidence showed that the bulk of the GRXE is produced
collectively by millions of weak X-ray sources 
mostly belonging to the old stellar population of the Galaxy --
cataclysmic variables (CVs) and coronally active stars in close
binaries (active binaries, or ABs)
\citep{mikej06,sazonov06,krivonos06,mikej_gc}. The X-ray 
luminosities of most of these sources are less than $10^{31}~{\rm
erg~s^{-1}}$, which makes the resolution of the GRXE into discrete sources a
challenging (although not completely infeasible) problem.

Recently, the cumulative X-ray emissivity of faint Galactic point
sources per unit stellar mass has been directly determined (by
integrating the X-ray luminosity function from $\sim 10^{27}$ to $\sim
10^{34}$~erg~s$^{-1}$) using an X-ray selected sample of sources
located in the vicinity of the Sun \citep{sazonov06}. However, given
the small size of this sample and possible pecularities
of the solar neighborhood, the resulting emissivity estimate may
differ from the average value for the Galactic stellar population by a
significant factor of $\sim 2$. It is very important to further reduce
this uncertainty in order to put a firm upper limit on any truly diffuse
contribution to the GRXE. This is particularly important for
understanding the physical processes taking place in the interstellar
medium. Unfortunately, the presently available data for the Milky Way do not
allow a considerable improvemement of the situation. 

One could hope to directly measure the cumulative X-ray emissivity of
stellar populations in other galaxies, provided they contain very
little hot interstellar gas and the contribution of bright LMXBs and
HMXBs can be completely resolved. Such a measurement, in particular,
would provide us important information about the cumulative soft X-ray
emission (below $\sim 2$~keV) of stellar sources, which is practically
inaccessible in our Galaxy due to the strong interstellar absorption
through the Milky Way.

\begin{figure*}[htb]
\hbox{
\includegraphics[width=0.39\textwidth]{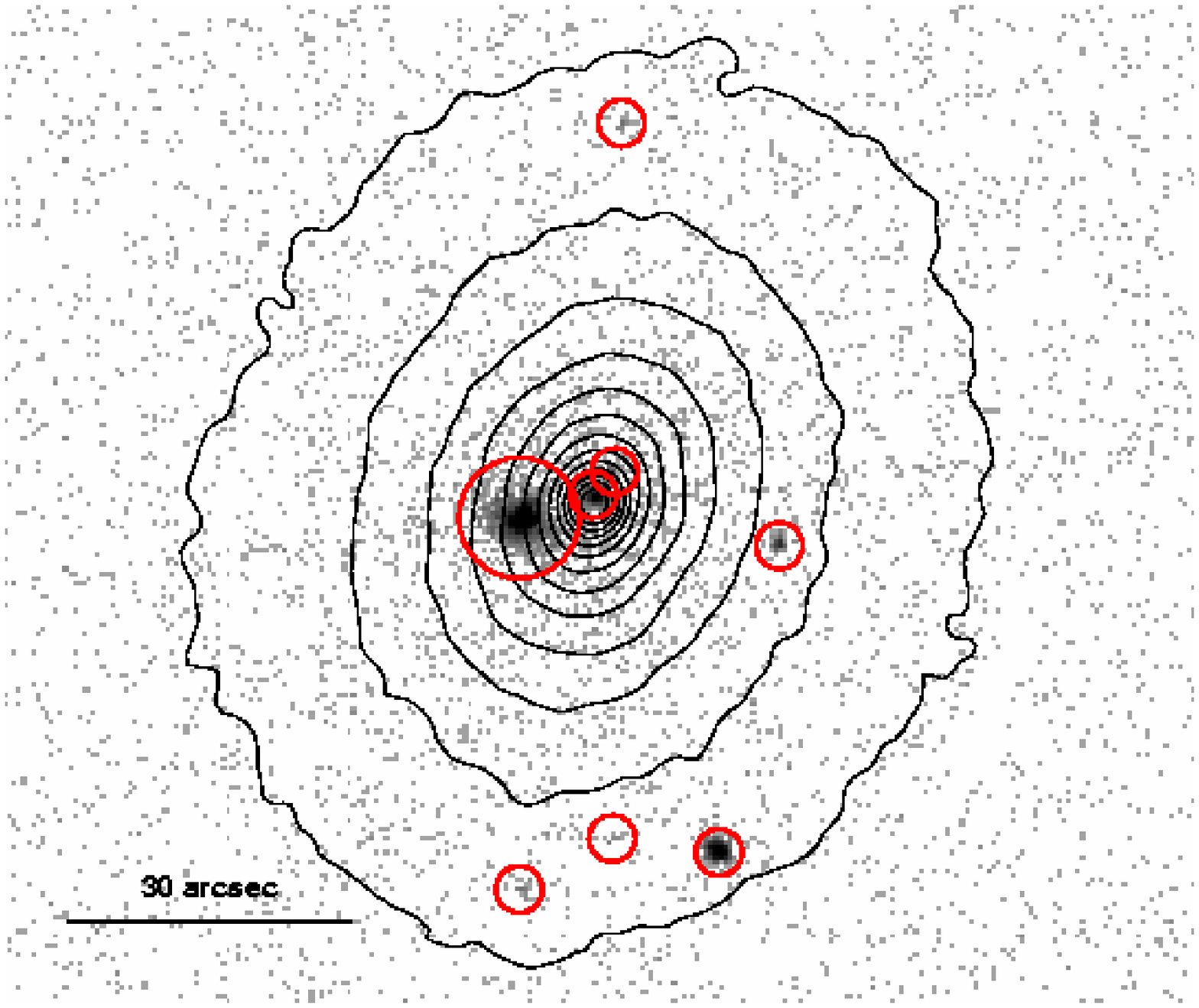}
\includegraphics[width=0.61\textwidth]{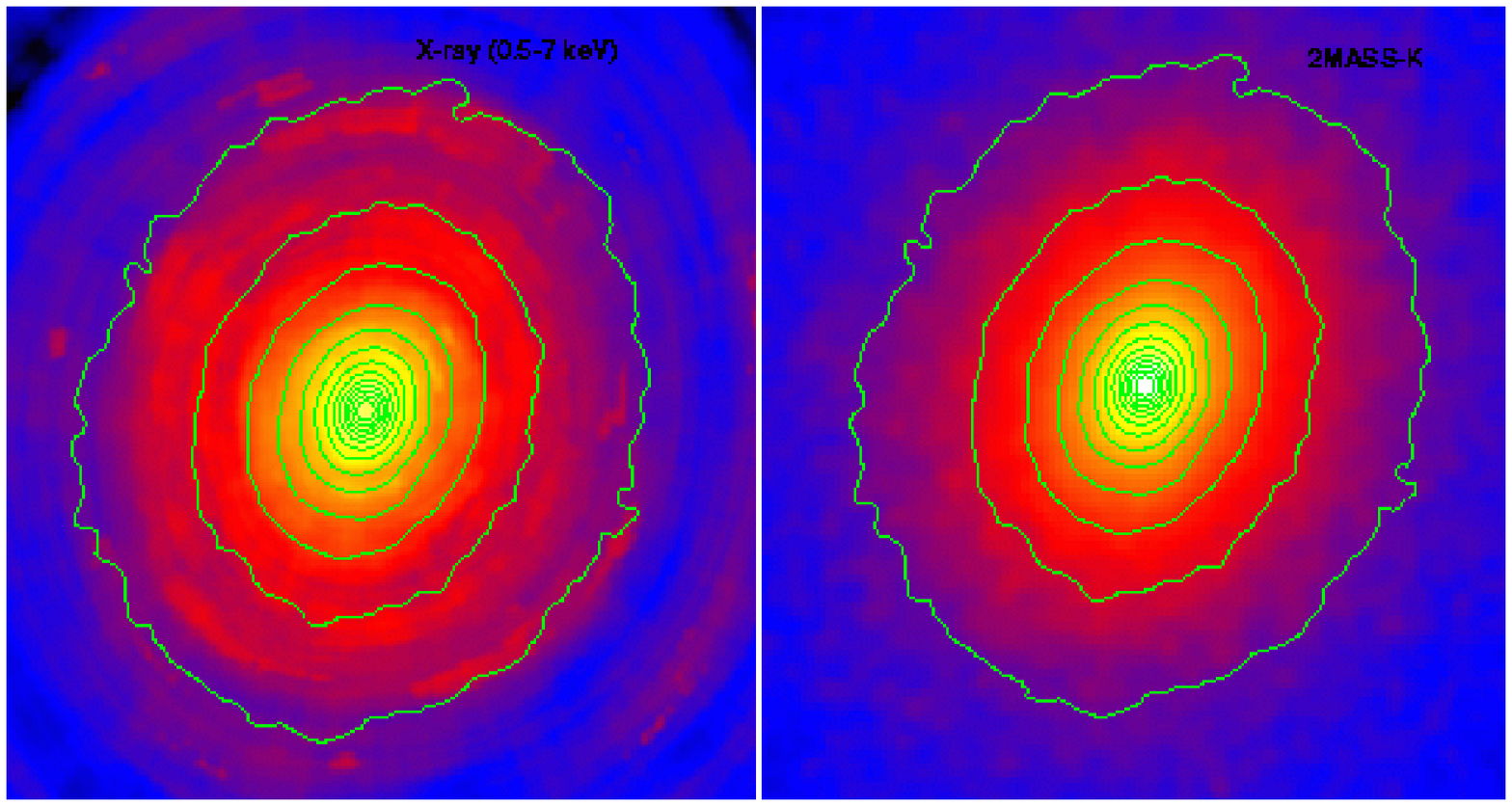}
}
\caption{{\it Left} -- raw {\it Chandra} image of M32 in the energy
band 0.3--7 keV. Contours denote isophotes of the galaxy in
near-infrared light. Circles show positions of detected sources. {\it
Center} -- adaptively smoothed {\it Chandra} image of M32 in the 0.5--7 keV
band with detected point sources removed. Contours denote isophotes
of the galaxy in near-infrared light (K-band, see right
panel). {\it Right} -- K-band image of the galaxy from
the 2MASS survey.
}
\label{image}
\end{figure*}

The best objects for performing this kind of measurement
are dwarf elliptical galaxies of the Local Group. Indeed, they i) lack
HMXBs and ii) they are close to us, which makes it possible to resolve
and subtract the contribution of virtually all LMXBs using the {\it
Chandra} telescope\footnote{The critical property here is the ability
of the telescope to resolve all sources with luminosities $L_{\rm
x}>10^{35}$ \lum, since at lower luminosities the contribution of
LMXBs becomes small. For a distance of 1 Mpc {\it Chandra} needs
$\sim$20 ksec to achieve this sensitivity.}. Also, their shallow
potential wells contain very little (if any) hot, X-ray emitting gas.

With these considerations in mind, we selected for a pilot study the
nearby -- 805 kpc \citep{mateo98} -- dwarf elliptical galaxy M32. This is
a well-known object, which has been extensively studied in practically
all wave-bands \citep{mateo98}. It has a total stellar mass of $\sim
10^{9}M_\odot$, contains very little cold gas \citep{welch01}, and
presents no evidence for recent star formation activity (REFS). The total
M32 X-ray luminosity (6--8$\times 10^{37}$ \lum) is dominated by a
single source, most likely a LMXB, M32 X-3
\cite[e.g.][]{loewenstein98,ho03}. Emission from the nucleus of the
galaxy and a supersoft X-ray source have also been detected
\citep{ho03}.

Previous analyses of {\it Chandra} observations of M82 revealed extended
X-ray emission  in addition to the detected point sources
\citep{ho03,fukazawa06}. \citet{ho03} proposed that this extended
emission is produced by hot interstellar gas in the galaxy.
 
We argue below that the unresolved X-ray emission in M32 is probably a
superposition of numerous faint point X-ray sources in the galaxy,
mostly CVs and ABs, and has essentially the same nature and properties
as the GRXE in the Milky Way.

The structure of the paper is as follows: in Sect. 2 we describe the
data used in the analysis, in Sect. 3 we present the obtained 
results on the morphology and spectrum of the unresolved emission in M32, 
in Sect. 4 we discuss possible origins of the unresolved emission, and
argue that the most plausible explanation of this emission 
is the cumulative contribution of weak unresolved sources of the old
stellar population in M32, at the end of Sect. 4 we discuss some
predictions for future observations of M32 and other gas-poor
elliptical galaxies, and we conclude in Sect. 5.

\section{Data analysis}
We used all publicly available {\em Chandra} observations of M32 in which 
the nucleus of the galaxy was far from the gaps between the
chips. Specifically, we analyzed observations with OBSIDs 313, 314, 1576, 1580,
1584, 2017, 2494, 2894, 2899, and 5690. The total exposure time of our
dataset is $\sim 215$ ksec.

The \emph{Chandra} data were reduced following a standard procedure
fully described in \cite{2005ApJ...628..655V}. The background for the
spectral analysis of the extended emission of M32 was obtained from a
$3^\prime$ by $1.5^\prime$ rectangle at $\sim 2^\prime$ to the south
of the center of the galaxy.

Point sources were detected using the wavelet decomposition package
 $wvdecomp$ of $ZHTOOLS$ \footnote{http://hea-www.harvard.edu/saord/zhtools/}. 
The source detection threshold for the analyzed dataset is
$\sim 10^{-16}$ \flux\ (in the energy band 0.5--7 keV), which
corresponds to a source luminosity $\sim 8\times10^{33}$ \lum\ (0.5--7
 keV) at the M32 distance.

\section{Results} 

\subsection{Morphology}

Figure~\ref{image} shows the 2' by 2' raw {\it Chandra} image of M32
in the 0.3--7 keV energy band. Within a radius of 50\arcsec\ around
the center of the galaxy we detected 8 point sources (including the
galactic nucleus), 3 of which (within 12\arcsec\ of the nucleus) were
previously reported by \citet{ho03} based on a shorter 50 ksec
exposure. The detected sources were excluded from the analysis of the
diffuse emission. In Fig.~\ref{image} (middle panel) an adaptively
smoothed X-ray image of M32 without point sources is shown with the
near-infrared isophotes from the 2MASS K-band image superposed.

In the optical and near-infrared bands, M32 has a nearly elliptical
shape with ellipticity 
$r_{\rm major}/r_{\rm minor}\sim 1.17$ \cite[see e.g.][]{kent87}.  In
order to construct radial profiles of the surface brightness of the
galaxy's extended (unresolved) X-ray emission, we extracted fluxes in
elliptical annuli defined by the above value of ellipticity and the
position angle derived from the isophotes of the galaxy in
near-infrared light (Fig.~\ref{image}).

\begin{figure}
\includegraphics[width=\columnwidth]{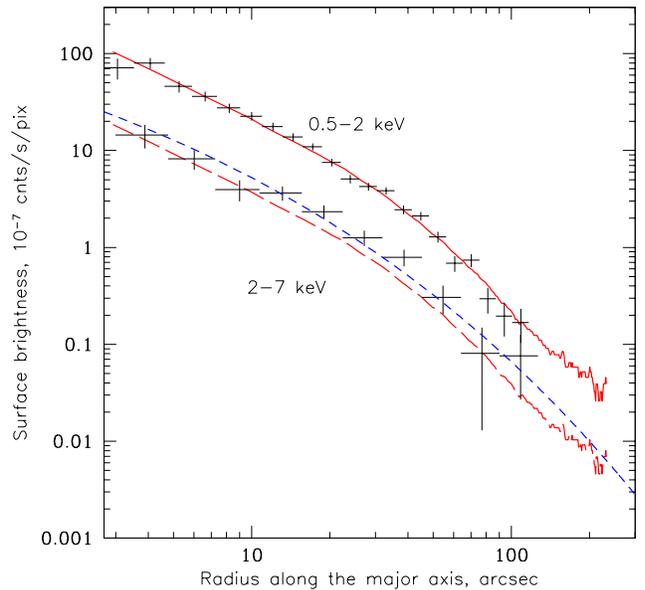}
\caption{Radial surface brightness profiles of M32 in two X-ray energy 
bands (crosses). The contribution of point sources with luminosities
$> 10^{34}$ \lum\  has been removed. The solid and long-dashed lines 
show the shape of the M32 profile in the near-infrared K-band
(arbitrary normalizations). The short-dashed line shows the de
Vaucouleurs law with core radius $r_{\rm c,major}=35\arcsec$ 
(arbitrary normalization).
}

\label{sb_profiles}
\end{figure}

The resulting radial surface-brightness profiles in two energy bands,
0.5--2~keV and 2--7~keV, are shown in Fig.~\ref{sb_profiles} together
with the near-infrared (K-band) light profile, derived from the 2MASS
image of M32 using the same elliptical annuli. Clearly the X-ray
intensity in both bands closely follows the near-infrared surface
brightness. The X-ray profile can be well fit by a de Vaucouleurs
law with effective radius $r_{\rm eff, major}=35\pm3\arcsec$. This
value is fully compatible with that obtained from optical imaging 
\cite[see e.g.][and also Fig.~\ref{sb_profiles} which compares the
  X-ray and optical light profiles]{kent87}. 

The constancy and value of the hardness ratio further support the
association of the unresolved X-ray emission with the stellar light
and its associated point sources. The hardness of the X-ray spectrum
(the ratio of the 2--7 and 0.5--2 keV fluxes) is consistent (within
errors) with being constant over radius from few to $\sim$100''
(Fig.~\ref{cum_hardness}). In addition, as Fig.~\ref{cum_hardness}
shows, the value of the hardness is charactestic of hard emission
(long dashed line) rather than that from soft (thermal) emission
(short dashed line).

\subsection{Spectrum}

\begin{figure}
\includegraphics[width=\columnwidth,bb=0 100 643 692,clip]{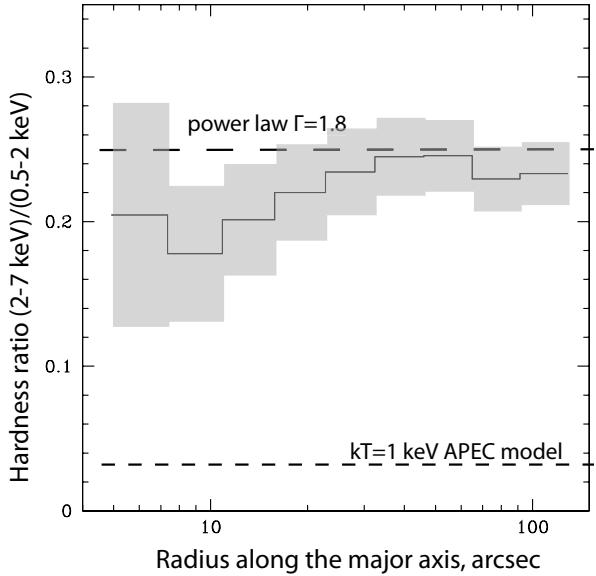}
\caption{Ratio of the X-ray fluxes in the energy bands 2--7~keV and
0.5--2~keV within a given radius from the center along the M32
major axis. The shaded area indicates the $1\sigma$ uncertainty in
this ratio. For comparison shown are the hardness ratios corresponding
to two representative spectral models: a power law and an optically
thin thermal emission.
}
\label{cum_hardness}
\end{figure}

\begin{figure}
\includegraphics[width=\columnwidth,bb=31 190 570 440,clip]{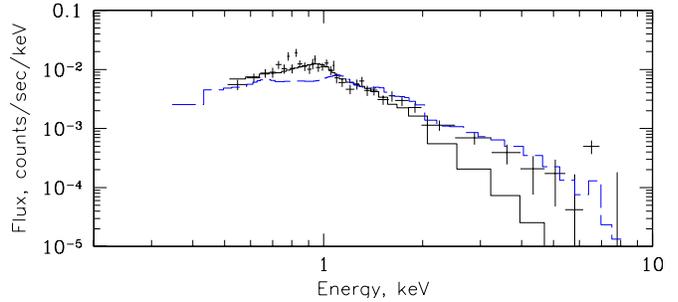}
\caption{Spectrum of the unresolved extended emission of M32. The solid
and dashed histograms represent optically thin plasma emission models
with $kT=1$ and 5~keV, respectively. The adopted abundance of elements
is 0.03 and 1.0 in the former and latter case, respectively. 
}
\label{spectrum_onemeka}
\end{figure}

In Fig.~\ref{spectrum_onemeka} we show the energy spectrum of the
unresolved extended X-ray emission. We accumulated this spectrum in
the elliptical region with major/minor axes of 62.5\arcsec/50\arcsec\
around the center of M32. This region contains 
$\sim70$\% of the total optical light (and hence stellar mass) of the
galaxy.

A single-temperature fit (APEC model, refs) to this spectrum in the 0.5--7
keV band is shown in Fig.~\ref{spectrum_onemeka} as the solid
histogram. The formal best-fit value of the temperature is $\sim$0.94
keV, while the best-fit abundance is extremely low, $\sim 0.04$ of
solar. These results are consistent with the values reported  by
\cite{ho03}. However, with the increased exposure time available now it
is clear that the $kT\sim 1$ keV plasma model fails to account for the
emission at energies $>$2--3 keV. Moreover, the hardness of the
unresolved X-ray emission of M32 was already evident from
Fig.~\ref{cum_hardness}. A much higher temperature $kT\sim 5$ keV is
needed to describe the data 
above $>2$ keV (blue histogram in Fig.~\ref{spectrum_onemeka}), but
this model cannot by itself explain the bulk of the soft X-ray
emission at energies below $\sim 1$~keV.  Thus, in terms of 
thermal plasma emission models, a multi-temperature plasma (with the
temperature distribution spanning a broad range) is necessary to explain
the observed spectrum in the 0.5--7 keV band.

\begin{figure}
\includegraphics[width=\columnwidth,bb=31 190 570 600,clip]{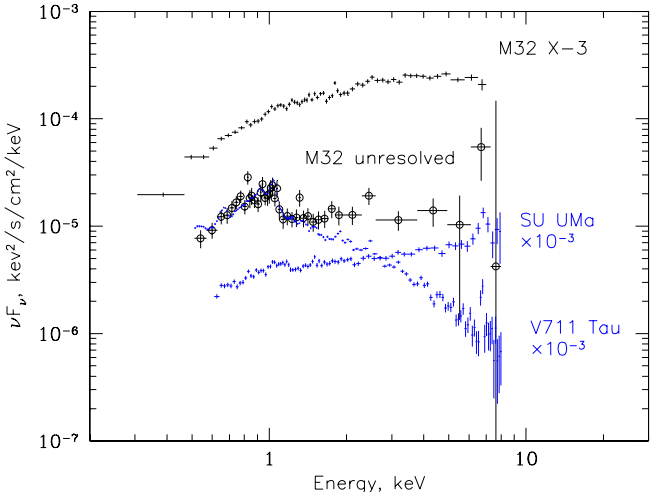}
\caption{Spectrum of the bright low-mass X-ray binary M32 X-3 and the
unfolded (through the power-law model with $\Gamma=2.0$) spectrum of 
the unresolved emission of M32. For illustration we also show the
arbitrarily scaled spectra (XMM-Newton/EPIC-MOS data) 
of a typical coronally active binary star
(V711 Tau) and a cataclysmic variable (SU Uma). Sources like these
probably make up the bulk of the unresolved emission in M32.}
\label{spectrum_lmxb}
\end{figure}

It is interesting to compare the spectrum of the unresolved emission
with the cumulative spectrum of the point sources detected by {\it
Chandra}. The central elliptical region of M32 introduced above is
dominated by the LMXB (probably containing a neutron star) M32
X-3. The time-averaged X-ray flux of this source is $F_{\rm 0.5-7 keV}
\sim 6.8\times 10^{-13}$ \flux, which translates into a luminosity
$L_{\rm 0.5-7 keV}\sim 5 \times 10^{37}$ \lum. In our Galaxy, LMXBs
with such luminosities are usually found in the so-called high
spectral state, when the optically thick accretion disk extends down
to the neutron-star surface. Therefore, most of the X-ray emission in
such systems is produced in the optically thick accretion disk and
boundary/spreading layer. The spectrum of M32 X-3 is shown in
Fig.~\ref{spectrum_lmxb}. Clearly it is very different from the
spectrum of unresolved emission.

\subsection{Luminosity}

Assuming a power-law shape of the spectrum with photon index $\Gamma=2.0$,
the 2--10 keV diffuse flux from the M32 central elliptical
region containing 70\% of the galaxy's optical light (see above) is 
$F_{\rm 2-10 keV}=(4.4\pm0.6)\times 10^{-14}$ \flux\ (flux is corrected
for the masked out circles around bright point sources). If we fit 
the spectral data, the best-fit slope parameter is 
$\Gamma=2.3\pm0.4$ and the 2--10 keV flux is $F_{\rm 2-10 keV}=(3.9\pm1.2)
\times10^{-14}$ \flux. Assuming a distance of 805 kpc \citep{mateo98},
we can then estimate the X-ray luminosity from the whole galaxy as
$L_{\rm 2-10 keV}=(4.9\pm0.7)\times 10^{36}$ \lum\ for $\Gamma=2.0$
and $L_{\rm 2-10 keV}=(4.3\pm1.3)\times 10^{36}$ \lum\ for $\Gamma=2.3$. 

We can estimate the near-infrared luminosity and
stellar mass of M32 from its total K-band magnitude of 5.095
(hereafter all optical and near-infrared 
characteristics of M32 are adopted from NED unless otherwise noted),
applying the distance modulus of 24.53, correcting for the interstellar 
extinction of 0.023 mag, and using the color-dependent K-band
mass-to-light ratio from \citet{bell03} for $B-V=0.88$. This 
results in a total K-band luminosity $L_{\rm K}=1.3\times
10^9~L_\odot$ and a total stellar mass
$M_\ast=1.0\times10^9~M_\odot$. Using these values, we can determine the 
emissivity of the diffuse X-ray component per unit stellar
luminosity and per unit stellar mass:
\[
\frac{L_{\rm 2-10~keV}}{L_{\rm K}} = 3.3\pm 1.0 \times10^{27} {\rm
erg~s^{-1}} L_\odot^{-1},
\]

\[
\frac{L_{\rm 2-10~keV}}{M_\ast}= 4.3\pm 1.3~(\pm 1.3)\times10^{27}
{\rm erg~s^{-1}} M_\odot^{-1}.
\]

The quoted uncertanties derive from the statistical errors 
on the measured X-ray flux. For the $L_{\rm 2-10~keV}/M_\ast$ ratio we have 
assumed that an additional uncertainty $\sim 30$\% might be associated
with the $L_{\rm K}$ to $M_\ast$ conversion (see
e.g. \citealt{bell03}). This systematic uncertainty is given in parentheses.

\section{Discussion}

Apart from the dominant X-ray emission from resolved point sources in
M32, we have clearly detected extended unresolved emission (thus
confirming the result of \citealt{ho03}). The data presented above
suggest that the cumulative emission of faint, stellar-type X-ray
sources in M32 provides the dominant contribution to this apparently
''diffuse'' component, similarly to the case of the Milky Way's ridge
emission. However, before discussing this scenario we first consider
other possibilities  to produce diffuse-like X-ray emission as
seen by {\it Chandra} in M32. 

\subsection{Diffuse emission from hot interstellar gas?}

Let us consider the hypothesis that the detected extended emission
originates in a hot diffuse gas pervading M32 and see whether it is
consistent with observations.  

Assuming a $\beta$-model for the spatial distribution of gas in M32,
we can deproject the observed X-ray surface brightness profile and
then estimate the total mass of the hot gas within the galaxy.

Approximating the observed surface-brightness profile $S(r)$ at
$5\arcsec<r<120\arcsec$ by the model (hereafter in this section
we assume for simplicity that the gas distribution is spherically
symmetric)

\[
S(r)=S(0) \left[ 1+\left({r\over{r_c}}\right)2\right]^{3\beta+1/2},
\]
yields a core radius $r_c\sim5$\arcsec\ and $\beta\sim0.34$.  
Using the emission measure ($\sim \int{N^2 dV}$) of hot (we adopted here 
$kT=1$ keV)
plasma as seen by {\it Chandra} within 50\arcsec\ radius around the
nucleus, we can estimate the mass of the hot gas in the galaxy. The
inferred integrated mass of hot gas in the central $50\arcsec$ is
$\sim 0.5\times10^{5} M_\odot$.

If we now assume that this gas is in hydrostatic equilibrium within
the potential well of the galaxy and has a single temperature, we can
estimate the total gravitating mass of the galaxy within $\sim
50$\arcsec:
$$
M_{\rm grav}(<r)={-kT_{\rm gas}\over{G\mu m_{\rm H}}}\left( {d \log \rho_{\rm gas} \over{d \log r}} \right).
$$
We thus find that the total mass of the galaxy within 50\arcsec\ is $
\sim 10^{10} M_\odot$ (assuming a $kT_{\rm gas}=1$ keV).
This value is much higher than the total gravitating mass of M32
(7--9.4$~10^8 M_\odot$)  measured from the velocity dispersion of 
planetary nebulae \citep{nolthenius86}. We have already seen that a single
temperature model does not provide a satisfactory description of the
X-ray spectrum and a multi-temperature plasma is needed with
temperatures ranging from $\sim$1 up to $\sim$5 keV. To make the mass
estimate from the hydrostatic equilibrium condition consistent with the
one based on planetary nebulae, one would need to assume that the gas
has temperature $\sim 0.1$ keV -- in strong contrast with our
spectral measurements.  Therefore, we conclude that the hot gas cannot
be in hydrostatic equilibrium within the potential well of M32.

If we instead assume that there is a constant supply of gas from the
stars and this gas is freely escaping the galaxy, we can estimate how
many old stars are needed to maintain $\sim 10^{5} M_\odot$ of hot
($\sim1$ keV) gas within $\sim50\arcsec$. According to different
estimates, typical mass ejection rates in elliptical galaxies are
1--3$\times 10^{-11} \, (L_{\rm V}/L_{\odot, \rm V}) \, M_\odot $ yr$^{-1}$
\citep[e.g.][]{faber76,padovani93,athey02}. Therefore, the stars in
the central 50\arcsec\ region of M32 ($L_{\rm V}\sim 10^8 L_\odot$) are 
expected to provide
$\dot{M}\sim (1-3)\times 10^{-3} M_\odot$ year$^{-1}$ of gas. We should compare
this value with the expected mass loss rate of hot gas due to its
outflow from the galaxy ($\sim 200$ pc in size) at approximately the
sound speed ($c_s\sim 500$ km~s$^{-1}$ for a 1 keV gas): $\dot{M}\sim
0.5\times10^{5}M_\odot / (200{\rm [pc]}/400 {\rm [km~s^{-1}]})\sim 0.1
M_\odot$ year$^{-1}$. We conclude that the scenario where the hot gas
is being constantly replenished by the stellar population is also very
unlikely.

\begin{table*}[htb]
\caption{X-ray emissivity of the old stellar populations (sources fainter 
than $\sim 10^{34}$ \lum\ in the 2--10~keV band) in M32 (measured by
{\it Chandra}) and in the Milky Way. 
}
\tabcolsep=0.5cm
\begin{center}
\begin{tabular}{c|c|c|c}
   & 0.5--2 keV & 2--7 keV & 2--10 keV\\ 
\hline
\hline
\multicolumn{4}{c}{M32$^{\rm a}$}\\
\hline
$L_{\rm x}/L_{\rm K}$, $10^{27}$ \lum\ $L_\odot^{-1}$&$4.1\pm0.6$&$2.7\pm0.8$&$3.3\pm1.0$\\
$L_{\rm x}/M_{*}$, $10^{27}$ \lum
$M_\odot^{-1}$&$5.4\pm0.7(\pm1.6)$&$3.5\pm1.0(\pm1.3)$&$4.3\pm1.3(\pm1.3)$\\
\hline
\multicolumn{4}{c}{Milky Way, GRXE$^{\rm b}$ \citep{mikej06}}\\
\hline
$L_{\rm x}/M_{*}$, $10^{27}$ \lum $M_\odot^{-1}$&$-$& $2.4\pm0.4(\pm1.0)$&$3.1\pm0.5 (\pm1.2)$\\
\hline
\multicolumn{4}{c}{Milky Way, solar vicinity$^{\rm c}$ \citep{sazonov06}}\\
\hline
$L_{\rm x}/M_{*}$, $10^{27}$ \lum $M_\odot^{-1}$ & $9\pm3$ & $2.4\pm
0.6$ & $3.1\pm 0.8$ \\ 
\hline
\end{tabular}
\end{center}
\begin{list}{}
\item $^{\rm a}$ -- the errors quoted in parentheses 
for the $L_{\rm x}/M_{*}$ ratio indicate the possible
uncertainty in the $L_{\rm K}$ to $M_\ast$ conversion, which is
probably $\sim 30$\%. 

\item $^{\rm b}$ -- the quoted values were derived assuming that all
of the GRXE originates from weak unresolved point sources; these
values were recalculated from the 3--20 keV band assuming a power-law
spectrum with $\Gamma=2.1$ \citep{mikej03}. The errors quoted
in parentheses indicate the uncertainty in the assumed mass of the
Galactic bulge \cite[see details in][]{mikej06} 

\item $^{\rm c}$ -- the quoted values exclude the
contribution of young stars, which is significant in
the solar neighborhood \citep{sazonov06}. Such sources are expected to
be almost completely absent in early-type galaxies like M32, while
their overall contribution to the GRXE is unknown. The uncertainty in 
the fraction of young stars mostly affects the lowest energy bin 
0.5--2.0 keV; the total cumulative emissivity of all sources (ABs,
young stars and CVs) near the Sun in this bin is $\sim 27\times
10^{27}$ \lum\ $M^{-1}_\odot$. The 0.5--2~keV emissivity was estimated
by multiplying 
the original 0.1--2.4~keV value by 0.7 (see \citealt{sazonov06} for
details). The 2--7~keV emissivity was 
converted from the 2--10~keV band assuming a power-law spectrum with
$\Gamma=2.1$, as for the GRXE. 

\end{list}

\label{tab:emis}
\end{table*}

\begin{figure}
\includegraphics[width=\columnwidth,bb=10 100 600 700,clip]{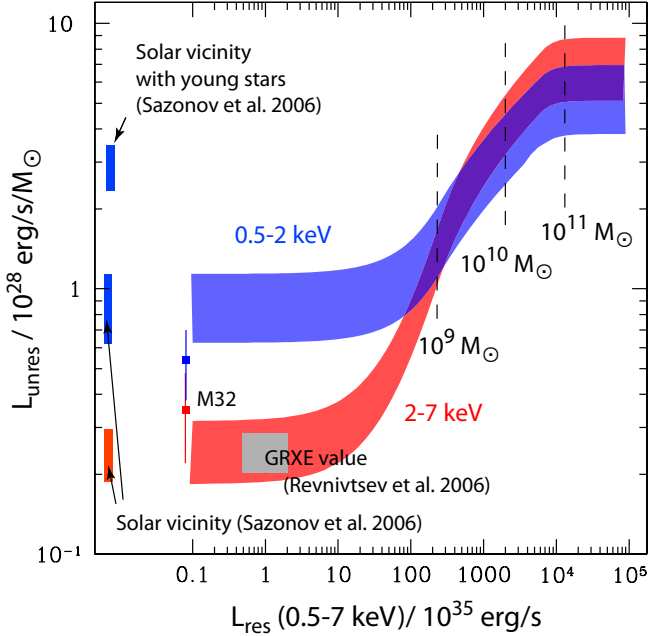}
\caption{Cumulative emissivity of unresolved discrete sources as a
function of the source detection threshold in a given observation. The
shaded areas are derived for two energy bands by integrating the
luminosity function of 
LMXBs from \cite{gilfanov04} and adding the cumulative emissivity of
fainter X-ray emitting systems from \cite{sazonov06} (see also Table
\ref{tab:emis}). The widths of these regions were fixed at 30\% to
emphasize the possible uncertainties in the stellar mass 
estimates involved in the measurements. Also shown are values obtained
from studies of the Galactic ridge X-ray emission (Revnivstev et al.,
2006), Solar vicinity (Sazonov et al., 2006) and M32 galaxy (this
work). The two lower emissivity estimates from \cite{sazonov06} exclude 
the contribution of young stars. Dashed lines mark several masses of
galaxies for which on average 
one LMXB with a 0.5--7 keV luminosity higher than a given value is expected
to be found.}
\label{resolved}
\end{figure}

\subsection{Emission from unresolved sources -- the old stellar population}
We have demonstrated that the hot insterstellar gas model clearly
fails to account for the observed diffuse emission of M32. We are now
turning to the hypothesis that the 
bulk of the diffuse emission is due to weak sources.  Given the
excellent sensitivity of the {\it Chandra} dataset for M32, we were able
to resolve all X-ray sources with luminosities higher than $L_{\rm
x}\sim 8\times 10^{33}$ \lum. However, still lower luminosity sources remain
undetected and could provide the unresolved X-ray flux should their
space density be sufficiently high.

In this regard we note that stellar-type X-ray sources with
luminosities below $10^{34}$ \lum\ are abundant in our Galaxy, in
particular in its old stellar population, the majority of these
sources being cataclysmic variables (CVs) and active binaries (ABs)
\citep[see e.g.][]{vaiana81,sazonov06}. Therefore, for a relatively old galaxy 
with very low star formation rate, like M32, one should also expect a
considerable X-ray luminosity from the numerous CVs and ABs. From the
luminosity and other properties of the GRXE 
\cite[e.g.][]{mikej06} and from direct measurements of the luminosity
function of sources in the solar neighborhood \citep{sazonov06}, the
combined 2--10~keV emissivity of CVs and ABs 
has been estimated as $L_{\rm x}/M_\ast \sim$ 2--4.5$\times10^{27}$
\lum $M_\odot$. The value $(4.3\pm1.3\pm 1.3)\times 10^{27}$ \lum $M_\odot$
obtained from observations of M32 falls exactly in this
range. This, together with the similarity of the X-ray and
near-infrared surface brightness maps, strongly suggests that in
M32 we are dealing with essentially the same phenomenon as the GRXE in
our Galaxy, namely the collective X-ray emission from millions of
faint, stellar-type sources.

Within this scenario, one should expect the spectrum of the M32
diffuse X-ray emission to be complex. Indeed, 
although we expect all of the unresolved emission to originate in
optically-thin thermal plasmas, these, in contrast to e.g. the
case of intracluster gas in galaxy clusters, should be characterized
by a broad range of temperatures. Indeed, active stellar coronae typically 
have temperatures in the range 0.1--3 keV, while the X-ray emitting regions in
CVs have much higher temperatures up to 20--30 keV. To
illustrate this point we show in Fig.~\ref{spectrum_lmxb} typical
spectra of an AB (V711 Tau) and a CV (SU
UMa). It is evident that a combination of such  
spectra could well resemble the observed spectrum of the M32 X-ray
halo. Specifically, the 0.8--1~keV peak would be mostly due to ABs,
while the harder X-ray tail due to both ABs and CVs. We do not 
attempt to carry out a more quantitative modelling of the 
observed spectrum along these lines, since there are significant
uncertainties in our understanding of the dependence of 
stellar coronal temperatures on luminosity, stellar type and age, as
well as of the soft X-ray spectra of different subclasses of CVs.

Table~\ref{tab:emis} summarizes several independent estimates of the
X-ray emissivity of the old stellar population per unit stellar mass
in different energy bands: 0.5--2~keV, 2--7~keV, and 2--10~keV. These
estimates include the current one based on M32, for GRXE and for the
solar neighborhood. 

We note that despite the excellent agreement at
energies above 2~keV, there is a hint that the local (near the
Sun) soft X-ray (0.5--2~keV) emissivity, even excluding young (age
$\ll 1$~Gyr) stars\footnote{Inclusion of young stars would increase
the local emissivity even further, up to $L_{\rm 0.5-2  
keV}/M \sim 27\times10^{27}$\lum\ $M^{-1}_\odot$.}, is somewhat higher
than the value obtained for M32. Since the soft X-ray emissivity is
dominated by ABs, this difference may indicate an intrinsic
difference in the old stellar populations of M32 and our local
environment.  We might also anticipate that galaxies with a star
formation history similar to that of our Galaxy should be charactered
by a cumulative emissivity closer to that obtained near the Sun than
to that obtained for M32. 

It is well-known that LMXBs can make an important
contribution to the X-ray flux of an elliptical galaxy, especially for
a gas poor one. If the LMXBs are bright enough to be directly
resolved, then their contribution can be accounted for and in the
remaining unresolved flux the role of weaker objects will be
higher. The dependence of the cumulative X-ray emissivity of the
unresolved stellar population as a function of sensitivity (minimum 
luminosity 
of resolved sources) is shown in Fig.~\ref{resolved}. The shaded areas
are the results of integration of the luminosity function of LMXBs determined
by \cite{gilfanov04} with the added cumulative emissivity of ABs and
CVs adopted from \cite{sazonov06} (see also Table
\ref{tab:emis}). From this figure it is clear that if sources with
luminosities $L_{\rm 0.5-7 keV}\ge 10^{35-36}$ \lum\ can be resolved, the
unresolved emission will be dominated by CVs and ABs. It is
worth emphasizing that in the 0.5-2 keV energy band the contrast
between the cumulative contributions of LMXBs and CVs/ABs is much
smaller than in the harder 2--10 keV band. Moreover, bright LMXBs
are rare objects and the probability of finding them in a small galaxy
is low. In Fig.~\ref{resolved} the dashed lines mark the masses of
galaxies in which on average one LMXB with a 0.5--7 keV luminosity
higher than a given value is expected to be found. Thus for a typical
gas poor galaxy with mass less then $10^{10} M_\odot$, which will
typically miss very bright sources with luminosities $>10^{38}$ \lum,
the contribution of CVs/ABs to the total X-ray emissivity of the
galaxy can be $\sim$ 30\% or higher.

While the 3--20 keV cumulative spectrum of the old
stellar population in our Galaxy has been reliably measured (Revnivtsev et
al., 2006), it is very difficult to extend this analysis to energies
below $\sim$3 keV because of the strong interstellar
photoabsorption in the Galaxy. M32 data do not suffer from
such strong absorption. Therefore, assuming that the X-ray
emissivities of the old stellar populations in our Galaxy and in M32
are similar, we can combine all of available data to build the
cumulative spectrum of the stellar populations in a broad energy
range. 

\begin{figure}
\includegraphics[width=\columnwidth,bb=0 95 605 700,clip]{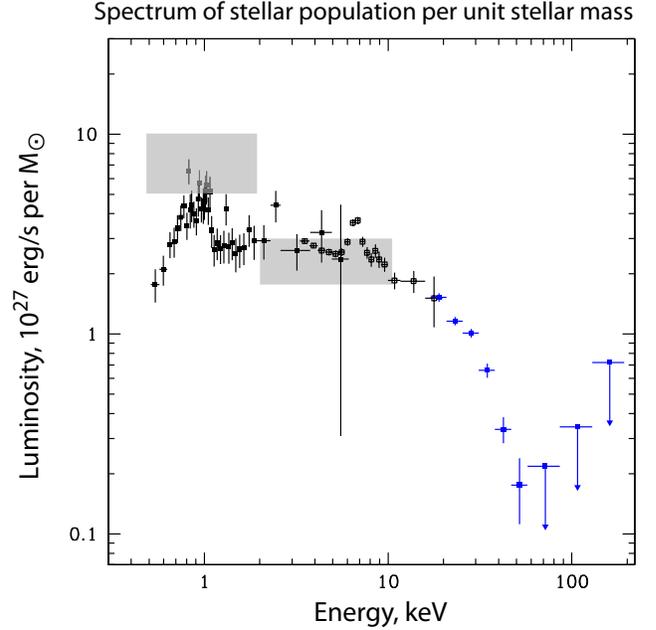}
\caption{Broad-band spectrum of the weak X-ray source population of
a typical old galaxy, compiled from data for the GRXE (3--20~keV -- {\sl
RXTE/PCA}, \citealt{mikej06}, 20--200~keV -- {\sl INTEGRAL/IBIS},
\citealt{krivonos06}) and for M32 (0.5--7~keV -- {\sl Chandra}, this
work). For M32 a mass of $M=10^{9} M_\odot$ is assumed. No fiducial
factors were applied to build this spectrum over almost three orders 
of magnitude in energy. Gray rectangles represent the cumulative
emissivities of X-ray sources, excluding young stars, in the Solar
neighborhood \citep{sazonov06} (see Table~\ref{tab:emis}).}
\label{broad}
\end{figure}

For this purpose we use our {\it Chandra} results on M32 and
{\sl RXTE/PCA} \citep{mikej06} and {\sl INTEGRAL/IBIS}
\citep{krivonos06} on the Milky Way. The resulting broad band spectrum
of the cumulative emission of weak X-ray sources is shown in Fig.~\ref{broad}.
In the high-energy end of this spectrum ($>$10--20 keV), intermediate
polars dominate \citep{mikej06, krivonos06}, while at low energies
coronally active stars contribute most significantly. The gray
rectangles show data obtained for the Solar vicinity (Sazonov et
al., 2006). Note that the 0.5--2 keV emissivity calculated for the Solar
vicinity is higher than the corresponding value for M32, even though
the former already excludes the contribution of young 
stars. As discussed above,
this tentative difference in the emissivities in the 0.5--2 keV band
may reflect the younger age of stars near the Sun compared
to M32.

\subsection{Predictions and future work}
An obvious prediction from our study is that any elliptical galaxy
with an optical luminosity $L_{\rm opt}$ should have a minumum level
of X-ray luminosity $L_{\rm x} \sim$ 3--5 $\times
10^{39}(M_\ast/10^{12}M_\odot)$ \lum\ even if it is very gas poor and
all bright LMXBs have been resolved. 

\paragraph{6.7 keV emission line } 
We have demonstrated that the unresolved X-ray halo in M32 can be
satisfactorily explained as the cumulative emission from CVs and
ABs, similar to the GRXE. In this regard we note
that a salient characteristic feature of the GRXE is the strong
complex of emission lines at 6.4 keV, 6.7 keV, and 6.9 keV, due to
highly ionized iron (6.7 and 6.9 keV lines) and fluorescenece of iron
in a relatively cold medium (6.4 keV). One could then expect M32 to
exhibit similar strong emission lines (with a combined equivalent
width of up to $\sim$600 eV, see e.g. \citealt{koyama86}). The
current statistics from {\it Chandra} do not allow us to claim the
presence of these lines, although we do see an indication of iron
emission. In order to securely detect the 6.7 keV line with {\it Chandra}, one
would need $\sim$1 Msec worth of observations.

\paragraph{Absence of ``X-ray cavities''}
If our hypothesis for the compact-source nature of the unresolved
emission in M32 is correct, one could anticipate there should be no
cavities produced in this X-ray halo by the plasma pressure of
radio-bubbles, as is often observed in elliptical galaxies or clusters
\cite[e.g.][]{bohringer95,churazov01,birzan04}. Unlike trully diffuse gas,
relativistic plasma should have no impact on the X-ray emission
associated with the stars (the same is true for LMXBs). While in M32
there is no evidence for any powerful relativistic plasma, other
galaxies could provide an opportunity to test this prediction. We note
however that the surface brightness of the X-ray emission due to old
stars is rather low and it may be problematic to detect the old
stellar component unambiguosly in
gas rich and/or more distant galaxies.

\section{Conclusions}

We have shown that the unresolved X-ray halo in the M32 dwarf
elliptical galaxy can be best explained by GRXE-like emission,
i.e. cumulative emission from cataclysmic variables and coronally
active stars. We have combined the spectra of the M32 diffuse emission
and of the GRXE to obtain a broad-band X-ray spectrum of the old
stellar population in our and other galaxies. We also derived
predictions of the unresolved X-ray emissivity (X-ray luminosity per
solar mass) of faint, stellar X-ray sources, in both soft (0.5--2~keV)
and hard (2--7~keV) bands, as a function of observational
sensitivity (minimum luminosity of resolved sources)
(Fig.~\ref{resolved}) 
which is critical for understanding the unresolved emission from galaxies.

\begin{acknowledgements}

This research made use of data obtained from the High Energy Astrophysics
Science Archive Research Center Online Service, provided by the
NASA/Goddard Space Flight Center. This publication makes use of data 
products from the Two Micron All Sky Survey, which is a joint project 
of the University of Massachusetts and the Infrared Processing and 
Analysis Center/California Institute of Technology, funded by the 
National Aeronautics and Space Administration and the National 
Science Foundation. This work was supported by DFG-Schwerpunktprogramme 
(SPP 1177)

\end{acknowledgements}

\end{document}